\documentclass[preprint,showpacs,prb,amsmath,amssymb,floatfix]{revtex4}
\usepackage{amsmath,amssymb}
\usepackage{bm}
\usepackage{epsfig}
\usepackage{psfrag}

\begin{document}

\title{Damping of zero sound in  Luttinger liquids}

\author{Peyman Pirooznia and Peter Kopietz}
  
\affiliation{Institut f\"{u}r Theoretische Physik, Universit\"{a}t
  Frankfurt,  Max-von-Laue Strasse 1, 60438 Frankfurt, Germany}

\date{June 20, 2007}

 \begin{abstract}
We calculate the damping $\gamma_q$ of collective density oscillations
(zero sound) in a 
one-dimensional  Fermi gas
with dimensionless forward scattering interaction 
$F$ and quadratic energy dispersion
$ k^2 / 2 m$ at zero temperature. 
Using standard many-body perturbation theory,
we obtain $\gamma_q$ from the expansion 
of the
inverse irreducible polarization to first order
in the effective screened (RPA) interaction.  
For wave-vectors $| q| /k_F  \ll  F$ (where $k_F = m v_F$ is the Fermi wave-vector)
we find to leading order $\gamma_q \propto | q |^3 /(v_F m^2)$.
On the other hand, for $F \ll | q| /k_F $ most of the spectral weight
is carried by the particle-hole continuum, which is
distributed over a frequency interval of the order of $q^2/m$.
We also show that zero sound damping 
leads to a finite maximum proportional to
$ |k - k_F |^{ -2 + 2 \eta}$
of the charge peak in the single-particle
spectral function, where $\eta$ is the anomalous 
dimension.  Our prediction agrees with photoemission data for
the blue bronze ${\rm K}_{0.3}{\rm MoO}_3$. 
We  comment on other recent calculations of $\gamma_q$.

\end{abstract}


\maketitle

\section{Introduction}
The normal metallic state of interacting electrons in
one spatial dimension has rather exotic properties, which are
summarized under the name Luttinger liquid behavior: the absence of a discontinuity
in the momentum distribution function at the Fermi surface,
a vanishing density of states at the Fermi energy, and an unusual 
line-shape of the single-particle spectral function,
characterized by algebraic singularities and separate peaks
for spin- and charge excitations \cite{Haldane81,Giamarchi04}.
To discover these features und study them quantitatively, it has been extremely
useful to have an exactly solvable effective low energy model
for interacting one-dimensional clean metals, the so-called Tomonaga-Luttinger model (TLM)~\cite{Tomonaga50, Luttinger63}. The exact solubility of the TLM relies on two crucial 
assumptions: the linearization of the energy dispersion for momenta close to
the two Fermi momenta $\pm k_F$, and the restriction to two-body scattering
processes involving only momentum transfers small compared with
$k_F$ (forward scattering).
The single-particle Green function is then most conveniently obtained via
bosonization \cite{Haldane81,Giamarchi04}. Alternatively, 
the Ward identity associated with the separate number 
conservation at each Fermi point can be used to
derive a closed  equation  for 
the single-particle Green function \cite{Dzyaloshinskii74}.
Within the Ward-identity approach, it is also straightforward to
show that collective density oscillations 
can propagate without damping in the TLM, so that the
random-phase approximation (RPA) for the density-density correlation function
$ \Pi ( q , \omega )$ is exact. As a consequence
the dynamic structure factor $S ( q , \omega ) = \pi^{-1}
 {\rm Im } \Pi ( q , \omega + i 0 )$
of the TLM exhibits a sharp $\delta$-function peak.
Focusing for simplicity on the spinless TLM with
forward scattering interactions $g_2 = g_4 = f_0$ in ``g-ology''-notation \cite{Haldane81,Giamarchi04}, the dynamic structure factor is
 \begin{eqnarray}
  S_{\rm TLM} (q ,\omega)  = Z_q \delta ( \omega - \omega_q )
 \; ,
 \label{eq:STLM}
 \end{eqnarray}
where  $Z_q=  | q | / ( 2 \pi \sqrt{ 1 + F} )$ and 
$\omega_q = v_c | q |$.
Here $F = f_0 / ( \pi v_F )$ is the relevant dimensionless interaction,
 $v_c = v_F \sqrt{1 + F}$
is the velocity of the collective charge excitations (zero sound, abbreviated by
ZS from now on), and
$v_F$ is the Fermi velocity.

The simple result (\ref{eq:STLM}) is a consequence
of the approximations inherent in the definition of the TLM: the
linearization of the energy dispersion and the restriction to
forward scattering interactions. In more realistic models, we expect
that the ZS mode acquires a finite width. 
How does the line-shape of 
$S ( q , \omega ) $ change if we do not linearize the energy dispersion?
Since the RPA is exact for linear energy dispersion,
it is reasonable to use the RPA as a starting point for
quadratic dispersion and to try to calculate the corrections to the
RPA perturbatively.
At the first sight it seems that 
this problem can be solved within the usual bosonization approach
by treating the effective boson-interactions due to the band curvature
within conventional many-body perturbation 
theory for the bosonized problem\cite{Haldane81,Kopietz96,Kopietz97}.
However, for frequencies close to
$\omega_q $ this seems not to be possible, and infinite re-summations are
necessary \cite{Teber06,Aristov07}.
In fact, there are conflicting
results for the $q$-dependence of the damping $\gamma_q$ of the
ZS mode of one-dimensional fermions in the literature: 
while Capurro {\it{et al.}} \cite{Capurro02} 
found $\gamma_q \propto q^3$,
Samokhin \cite{Samokhin98},
Pustilnik and co-authors \cite{Pustilnik03,Pustilnik06,Pustilnik06b} and Pereira 
{\it{et al.}} \cite{Pereira06} obtained $\gamma_q \propto q^2$.
On the other hand,  Teber \cite{Teber05}  recently showed
that the attenuation rate of an acoustic mode embedded in the
two-pair continuum of the dynamic structure factor
scales as $q^3$, which is consistent with 
$\gamma_q \propto q^3$.

Let us briefly discuss the recent works  by 
 Pustilnik {\it{et al.}} \cite{Pustilnik06,Pustilnik06b}
and by Pereira {\it{et al.}} \cite{Pereira06},  both of which found
that the damping scales as $q^2$.
In Ref. [\onlinecite{Pereira06}] it has been shown that
the width $\gamma_q$ of the peak of the longitudinal structure factor
of the XXZ spin-chain is proportional to $q^2$.
This is not necessarily in conflict with  
$\gamma_q \propto q^3$ for the TLM with band curvature,
because the XXZ-chain is equivalent to a system of
spinless fermions on a lattice. In contrast to the forward scattering processes
of the TLM, the interaction in this model has also 
scattering processes involving momentum transfers of the order of
$k_F$. 
Although in the Luttinger liquid regime of the XXZ-chain
these processes are irrelevant in the renormalization group sense,
their effect on non-universal quantities like
$\gamma_q$ might be essential.
A similar argument applies also to Ref.~[\onlinecite{Pustilnik06b}],
where the dynamic structure factor of the
Calogero-Sutherland  model is shown  to differ from zero only
in a finite interval of frequencies of the width proportional to 
$ q^2 /m$.  The Fourier transform $f_q$ of the 
interaction in the Calogero-Sutherland model
is proportional to  $| q | /m$ for all  momentum transfers $q$, so that
scattering is suppressed for small $q$.
On the other hand, the RPA and our strategy of
calculating perturbative corrections to the RPA can only be expected to
be accurate  if the interaction $f_q$ is  strongest for small $q$.
We therefore believe that the dynamic structure factor of the
Calogero-Sutherland model~\cite{Pustilnik06b}
does not represent the generic behavior 
of the dynamic structure factor of  one-dimensional Fermi systems
with dominant forward scattering, as given by the TLM with curvature.

As far as the calculation in Ref.~[\onlinecite{Pustilnik06}] is concerned,
it is based on an infinite re-summation
of the apparent leading  singularities in the weak coupling expansion,
using an analogy with the X-ray problem. We believe that this procedure does not
properly take into account  the asymptotic Ward-identity which
guarantees the cancellation of all singularities 
in the limit $1/m \rightarrow 0$.
Moreover, the renormalization of the real part of the
energy of the ZS mode, which might  be essential to obtain 
the correct damping \cite{Schoenhammer07,Pirooznia07}, has not been properly 
taken into account in Ref.~[\onlinecite{Pustilnik06}].

\begin{figure}[tb]    
\centering   
\psfrag{y}{$R_q$}
\psfrag {x}{$F$} 
  \vspace{4mm}
\includegraphics[scale=0.3]{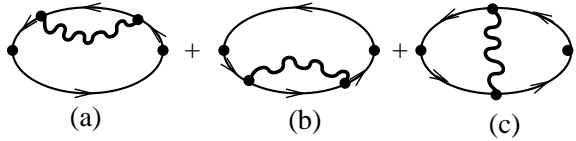} 
\caption{Leading interaction corrections to the
irreducible polarization in an expansion in powers of the
RPA interaction. Solid arrows are non-interacting Green functions and
wavy lines denote the RPA interaction. 
}
    \label{fig:feynman}
\end{figure}

Within the framework of diagrammatic many-body perturbation theory, 
the standard approach \cite{Holas79} to calculate corrections to the RPA is based
on the evaluation of the three Feynman diagrams shown in Fig.~\ref{fig:feynman},
which represent the leading corrections
to the irreducible polarization in an expansion in powers of the RPA interaction.
Surprisingly, these diagrams have never been evaluated for the TLM
with band curvature, which we shall do in this work.
It seems that, at least for small interactions,
this approach should be sufficient to estimate the
damping of the ZS mode.
We shall further comment on the accuracy of this 
approach in  Sec.~\ref{sec:conclusions}.

\section{RPA in one dimension with quadratic dispersion}

\begin{figure}[tb]    
   \centering
\psfrag {A}{$F=0$} 
\psfrag{B}{\footnotesize $F=0.006$}
\psfrag {C}{\footnotesize$F=0.04$} 
\psfrag {D}{\footnotesize$F=0.06$} 
\psfrag {E}{\footnotesize$F=0.1$} 
\psfrag{y}{\hspace{-8mm} $ \pi v_F S_{\rm RPA} ( q , \omega ) $}
\psfrag {x}{$\omega/v_{F}|q|$} 
  \vspace{4mm}
\includegraphics[scale=0.8]{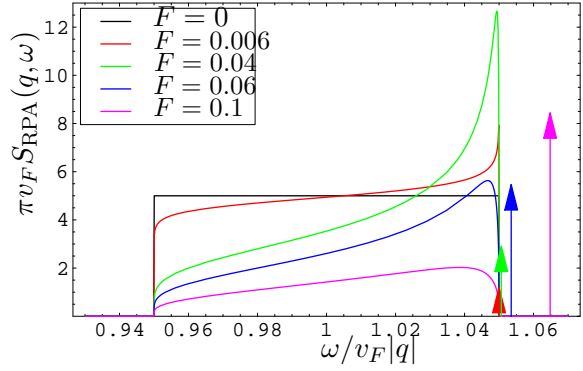} 
  \caption{RPA dynamic structure factor for 
quadratic energy dispersion,  $|q|/k_{F}=0.1$, and different values of the dimensionless
interaction $F$. The arrows denote the location of the 
$\delta$-peak associated with the ZS mode, the length of the arrows
being proportional to the relative weight $W_q$ of the ZS peak
in  the $f$-sum rule, see Eq.~(\ref{eq:Zqrpa}).
}
    \label{fig:Srpa}
\end{figure}

\begin{figure}[tb]    
\centering   
\psfrag{y}{$W_q$}
\psfrag {x}{$F$} 
  \vspace{4mm}
\includegraphics[scale=0.6]{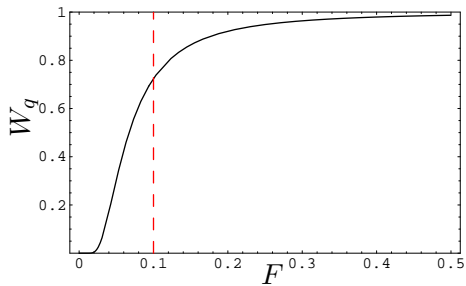} 
\caption{Weight $W_q$ 
of the ZS mode in the $f$-sum rule
in   RPA for $ q /  k_F = 0.1 $ (dashed line) as a function of $F$, see Eq.~(\ref{eq:Zqrpa}).
}
    \label{fig:fsum}
\end{figure}

It is instructive to consider first the density-density
correlation function for fermions with quadratic energy dispersion
$\epsilon_k = k^2 / 2 m$ 
within the RPA, where
\begin{equation}  
\Pi^{-1} _{\rm RPA}(Q) =  f_0  + \Pi^{-1}_{0}(Q) 
 \; .
\end{equation}
For convenience we use the Matsubara formalism
and collective labels
$Q = (  i \omega , q )$ for 
wave-vector $q$ and bosonic Matsubara frequency
$ i \omega $. 
To make contact with the usual distinction between left-moving and right-moving
fermions in the TLM, we write the non-interacting
polarization as
 $\Pi_{0}(Q)=\sum_{\alpha = \pm } \Pi_{0}^{\alpha} (Q)$, where
$\alpha = +$ refers to right-moving fermions (with velocity
$v_k = k/ m > 0$) while $\alpha =-$ denotes
left-moving fermions ($v_k < 0$).
At finite temperature $T$ for a system of length $L$,
\begin{eqnarray}
 \Pi_{0}^{\alpha} (Q) & = & - \frac{T}{L} \sum_{K} 
 \Theta^{\alpha} ( k )
G_0^{\alpha} (K) G_0^{\alpha} (K+Q)
\; , 
\label{eq:Pi0}
\end{eqnarray}
where  
$G_0^{\alpha } ( K ) = [ i \tilde{\omega} - \xi^{\alpha}_k ]^{-1}$ is the
free Matsubara Green function and
$\xi^{\alpha}_k = \epsilon_{ \alpha k_F + k} - \epsilon_{ k_F} = 
\alpha v_F k + k^2/2m$ is the free 
excitation energy.
 The label $K = ( i \tilde{\omega} , k )$ 
consists of fermionic Matsubara frequency $i \tilde{\omega}$ and momentum
$k$, which is measured relative to $\alpha k_F$.
The cutoff function $\Theta^{+} (k )$ restricts the
range of the $k$-integration to the interval $[ - k_F , \infty )$ associated with
right-moving fermions, while
 $\Theta^{-} (k )$ selects the interval $( - \infty , k_F]$
corresponding to left-moving fermions.
All degrees of freedom are taken into account in this way.
At $T=0$ the integrations in
Eq.~(\ref{eq:Pi0}) are easily carried out,
\begin{eqnarray}
\Pi_{0}^{\alpha}(Q)=\alpha\frac{m}{2\pi q}
\ln \left[ 
\frac{\alpha v_{F}q - i \omega + q^2/2m  }{\alpha v_{F}q - i \omega - q^2/2m }
 \right]
 \; ,
 \label{eq:Pi0res} 
\end{eqnarray}
so that  dynamic structure factor $S_{\rm RPA} ( q , \omega )
 = \pi^{-1} { \rm Im } \Pi_{\rm RPA} ( q , \omega + i 0 )$
can be calculated  analytically.
For any finite value of the interaction $S_{\rm RPA} ( q , \omega )$
consists of two contributions: a continuum part
due to partice-hole excitations, and a sharp
$\delta$-peak $S^{\rm zs}_{\rm RPA} ( q , \omega ) = Z_q
 \delta ( \omega - \omega_q )$ corresponding
to the undamped collective ZS mode, see Fig.~\ref{fig:Srpa}.
The
weight $Z_q$ and the dispersion $\omega_q$ of the
ZS mode are within RPA
 \begin{eqnarray}
Z_{q} &  = &  \frac{v_F q^2}{2 \pi  \omega_q }  W_q \; \; ,  \; \; W_q =
  ( \tilde{q}/F)^2  / \sinh^2({\tilde{q}}/{F})  \; , 
 \label{eq:Zqrpa}
\\
\omega_{q} & = & 
v_{F} | q| [1+ \tilde{q} \coth\left( \tilde{q}/F \right)+
 \left( \tilde{q} / 2 F \right)^2]^{1/2}
\; ,
\label{eq:omegaqrpa}
\end{eqnarray}
where $\tilde{q} = q / k_F $.
Obviously, the limits of vanishing band curvature
($ | \tilde q | = | q | / m v_F   \rightarrow 0$) and
vanishing interaction ($F \rightarrow 0$) do not commute:
for $| \tilde{q} | \gg F $ the weight $Z_q$
of the ZS mode is exponentially small, so that the
particle-hole part of the dynamic
structure is dominant.  For  $F \rightarrow 0$   the latter approaches  a
box-function centered at $\omega = v_F | q | $ of width
$q^2/m$ and height  $ m / ( 2 \pi | q | )$, see Fig.~\ref{fig:Srpa}.
In the opposite limit $ | \tilde{q} | \ll F$ the ZS peak
is dominant. To quantify this, note that the
dimensionless factor $W_q$ defined in Eq.~(\ref{eq:Zqrpa}) can be identified 
with the relative weight 
of the ZS peak in the $f$-sum rule
$
 \int_{0}^{\infty}\mathrm{d}\omega\,\omega \,S(q,\omega)= v_F q^2 / 2 \pi$.
From Fig.~\ref{fig:fsum} it is clear that
for $  | \tilde{ q }| \ll F$ the contribution of the particle-hole continuum
to the $f$-sum rule is indeed negligible (actually, it is of order $ ( \tilde{q} / F )^2  \ll 1$),
and  that the crossover between the particle-hole regime $ |\tilde{q} | \gg F$ and
the ZS regime $ | \tilde{q} | \ll F$  occurs at
$ | \tilde{ q  }| \approx F$.

The line-shape of the particle-hole continuum and its position relative to the
ZS mode are probably not correct within the RPA.
In fact, recently Sch\"{o}nhammer \cite{Schoenhammer07} has shown that
if one uses in the RPA bubbles  Hartree-Fock propagators instead of bare ones, 
then the energy of the ZS mode 
is {\it{smaller}} than the energy of the particle-hole continuum.
Moreover, multi-pair particle-hole excitations neglected within the RPA will
wash out the sharp thresholds predicted by the RPA and
generate some small spectral weight for all frequencies~\cite{Pines90}.
However, for sufficiently  small $q$ almost the entire spectral weight 
is carried by the ZS mode, so that in this limit we may neglect the
particle-hole continuum.

\section{Leading correction to the RPA}
 
Because for $  | q | /k_F \ll F$ the spectrum of the density fluctuations 
is  dominated by
the collective ZS mode, we expect that
the exact  structure factor  for frequencies $\omega$ close
to the exact ZS frequency $\omega_q$ 
can be approximated by a Lorentzian,
 \begin{equation}
 S ( q , \omega ) \approx \pi^{-1} {Z}_q  \gamma_q [  ( \omega -
 {\omega}_q )^2 + \gamma_q^2 ]^{-1}
 \; .
 \label{eq:Lorentz}
 \end{equation}
In terms of the 
irreducible polarization
$\Pi_{\ast} ( Q )$  defined via
$\Pi^{-1} ( Q ) = f_0 + \Pi_{\ast}^{-1} (Q )$
the inverse weight of the ZS peak is given by
 \begin{equation}
 {Z}_q^{-1} =   f_0^2  {\rm Re} 
\Pi_{\ast}^{\prime} ( q , \omega_q ),
 \end{equation}
 and the damping is
 \begin{equation}
\gamma_q =  {\rm Im} \Pi_{\ast} ( q , {\omega}_q + i 0) /
  {\rm Re} 
\Pi_{\ast}^{\prime} ( q , \omega_q ), 
 \end{equation}
where
$\Pi_{\ast}^{\prime} ( q , \omega ) = \partial \Pi_{\ast} ( q , \omega ) / \partial
\omega$.

Since $\gamma_q =0$ within RPA, we need to go beyond RPA to
estimate $\gamma_q$.
This is usually done \cite{Holas79} by
expanding  the $\Pi_{\ast} ( Q )$ 
in powers of the RPA interaction
$f_{\rm RPA} ( Q ) = f_0 [  1 + f_0 \Pi_{0} ( Q ) ]^{-1}$.
Before presenting an explicit calculation, 
let us give a simple argument for the expected 
$q$-dependence of $\gamma_q$.
Within the functional bosonization
approach \cite{Kopietz97}, 
$\Pi^{\alpha} (Q )$
can be written as a
single-particle Green function 
of a real bosonic quantum field ${\rho}^{\alpha}_Q$
representing the density fluctuations,
\begin{equation}
 \Pi^{\alpha} ( Q)  \propto  \int {\cal{D}} [ {\rho}^\alpha ] 
e^{ - S_{\rm eff} [ {\rho}^{\alpha} ] }
 {\rho}^{\alpha}_{ -Q}
   {\rho}^{\alpha}_{ Q} 
 \; .
 \label{eq:Seffrho}
 \end{equation}
For linear dispersion the effective action
$S_{\rm eff} [ {\rho}^{\alpha} ]$ is quadratic, so that 
Eq.~(\ref{eq:Seffrho}) yields the RPA result, which is exact 
for the TLM.
Non-linear terms in the energy dispersion renormalize
the quadratic part of $S_{\rm eff} [ {\rho}^{\alpha} ]$
and  give rise to cubic, quartic, and higher order retarded
interaction vertices \cite{Kopietz97}, which all
generate corrections to the RPA. 
We 
parameterize these corrections in terms of an irreducible self-energy
 $ {\Sigma}^{\alpha} ( Q )$. 
Assuming  that
$ {\Sigma}^{\alpha} ( Q )$ is analytic for small $Q$ (this assumption
relies on the cancellation of all singularities in the
symmetrized closed fermion loops~\cite{Pirooznia07}), and
taking into account that 
$ {\Sigma}^{\alpha} ( Q ) =  {\Sigma}^{\alpha} ( - Q )$
(which follows from the invariance of the
right-hand side of Eq.~(\ref{eq:Seffrho}) under $Q \rightarrow -Q$),
we conclude  that 
${\Sigma}^{\alpha} ( q , \omega_q + i 0 ) = c_0 + c_2 ( q / k_F )^2 + 
O ( q^4 )$, with real $c_0$ and complex $c_2$. 
But the 
relation between the density fields ${\rho}_Q^{\alpha}$ and the
Tomonaga-Luttinger bosons
whose damping $\gamma_q$ we are seeking involves an extra factor 
of $  | q |^{1/2}$ (see for example p.57 of Ref.~[\onlinecite{Kopietz97}]),  so that
we expect
$\gamma_q \propto | q | {\rm Im} {\Sigma}^{\alpha} ( q , \omega_q + i 0 ) 
\propto | q |^3$.

\begin{figure}[tb]    
   \centering
  \psfrag{z}{$q/k_{F}$}
\psfrag {y}{$\omega/v_{F} k_{F}$} 
 \psfrag{x}{$q/k_{F}$}  
\psfrag{a}{a}  
\psfrag{b}{b}  
\vspace{4mm}
\includegraphics[scale=0.60]{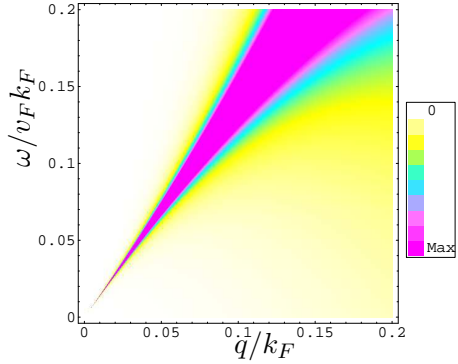}    
\caption{Dynamic structure factor
in Lorentzian approximation for $F =1$, see
Eqs.~(\ref{eq:Zqrpa},\ref{eq:omegaqrpa},\ref{eq:Lorentz}) and (\ref{eq:dampres}).
}
\label{fig:contour}
  \end{figure}

We now confirm this result by explicitly calculating
$\Sigma^{\alpha} ( Q )$ to first order in the RPA interaction. 
The diagrams contributing to $\Pi_{\ast} ( Q )$ to this order
are shown in Fig.~\ref{fig:feynman}.
Writing again $\Pi_{\ast} ( Q ) = \sum_{\alpha} \Pi^{\alpha}_{\ast} ( Q )$
we have 
$\Pi^{\alpha}_{\ast} ( Q ) \approx \Pi^{\alpha}_{0} ( Q )
 + \Pi^{\alpha}_{1} ( Q )$, with
 $\Pi^{\alpha}_{1} ( Q )
 = \Pi^{\alpha}_{1s} ( Q ) + \Pi^{\alpha}_{1v} ( Q )$.
The sum of the self-energy corrections
shown in Fig.~\ref{fig:feynman} (a,b) is
\begin{eqnarray} 
\Pi_{1s}^{\alpha} ( Q ) & = & \frac{T^2}{ L^2}\sum_{K,Q'}
 \Theta^{\alpha} ( k )
f_{\rm RPA} ( Q^{\prime})
 [ G_{0}^{\alpha}(K) ]^2  
 \nonumber
 \\
 &   & \hspace{-10mm} \times G_{0}^{\alpha}(K+Q')
 [ G_{0}^{\alpha}(K+Q) +  G_{0}^{\alpha}(K-Q) ]
 \; ,
 \label{eq:Pi1sv1}
\end{eqnarray}
and the vertex correction in Fig.~\ref{fig:feynman} (c) is
\begin{eqnarray} 
\Pi_{1v}^{\alpha} ( Q ) & = & \frac{T^2}{ L^2}\sum_{K,Q'}
 \Theta^{\alpha} ( k ) f_{\rm RPA} ( Q^{\prime})
  G_{0}^{\alpha}(K)    G_{0}^{\alpha}(K+Q)
 \nonumber
 \\
 & & \times
  G_{0}^{\alpha}(K+Q^{\prime})  G_{0}^{\alpha}(K+Q + Q^{\prime}) 
 \; .
 \label{eq:Pi1vv1}
\end{eqnarray}
To regularize some of the $q^{\prime}$-integrations,
we introduce a  momentum transfer cutoff
$q_c \ll k_F $ which restricts the integration range to
$ | q^{\prime} | \leq q_c $.  The imaginary part of these expressions
is not sensitive to  $q_c$. To
evaluate Eqs.~(\ref{eq:Pi1sv1},\ref{eq:Pi1vv1}) in the limit $ T \rightarrow 0$
it is convenient to write \cite{Kopietz97}
 \begin{equation}
 f_{\rm RPA} ( q, i \omega ) = f_0 - f_0^2 \int_0^{\infty} \mathrm{d} \omega^{\prime}
 \frac{ 2 \omega^{\prime} S_{\rm RPA} ( q , \omega^{\prime} )  }{ \omega^{\prime \;2 } + \omega^2 }
 \; .
 \label{eq:fRPAS}
 \end{equation} 
The frequency integrations in Eqs.~(\ref{eq:Pi1sv1}) and (\ref{eq:Pi1vv1})
can then be carried out exactly. 
 The result is
 \begin{widetext}
  \begin{eqnarray} 
 \Pi_{1s}^{\alpha} ( q , i \omega ) & = &
 \frac{f_0}{L^2}\sum_{k,q'} \Theta^{\alpha} ( k )
 \theta(-\xi_{k+q'}^{\alpha})
  \bigg[\frac{\delta( \xi_{k}^{\alpha}) }{
 \xi_{k+q}^{\alpha}- i\omega}
 +\frac{{\rm sgn} ( \xi_{k}^{\alpha} )
  \theta(-  \xi_{k}^{\alpha}  \xi_{k+q}^{\alpha})      }{[\xi_{k}^{\alpha}-\xi_{k+q}^{\alpha}+i\omega ]^2}
 \bigg]
 \nonumber\\ 
 &&
 \hspace{-23mm} 
 -\frac{f_0^{2}}{L^2}
  \sum_{k,q'} \Theta^{\alpha} ( k )   \int\limits_{0}^{\infty}\mathrm{d}\omega^{\prime}
 \, \Bigg\{
 \frac{ S_{\rm RPA}(q',\omega^{\prime} )   }{\xi_{k}^{\alpha}-\xi_{k+q}^{\alpha}+i\omega}
 \bigg[
 \frac{  {\rm sgn}( \xi_{k+q'}^{\alpha} )  \delta( \xi_{k}^{\alpha}) }{
 | \xi_{k+q'}^{\alpha} | +\omega^{\prime} }
 +
 \frac{ {\rm sgn} ( \xi_{k}^{\alpha} )    \theta(- \xi^{\alpha}_k \xi_{k+q'}^{\alpha})}{
 [ | \xi_{k}^{\alpha} | + | \xi_{k+q'}^{\alpha} | +\omega^{\prime}]^2} 
 \bigg]
 \nonumber
 \\ 
 && \hspace{-22mm} + \frac{  S_{\rm RPA}(q',\omega^{\prime} )   }{[\xi_{k}^{\alpha}-\xi_{k+q}^{\alpha}+i\omega]^2}\,
 \bigg[
 \frac{
 \theta(-\xi_{k}^{\alpha}\xi_{k+q'}^{\alpha}   )}{
 | \xi_{k}^{\alpha} | + | \xi_{k+q'}^{\alpha}| + \omega^{\prime}}
 - 
 \frac{\theta(- \xi_{k+q}^{\alpha} \xi_{k+q^{\prime}}^{\alpha})}{ | \xi_{k+q}^{\alpha}| +| \xi_{k+q'}^{\alpha}| +\omega^{\prime} + 
 i \omega \; {\rm sgn} ( \xi^{\alpha}_{k + q^{\prime}}) }
 \bigg]\Bigg\}
 \nonumber
 \\
 & + & 
  [  (q,i\omega )\to(-q,-i\omega ) ]
  \; ,
  \label{eq:Pi1s}
 \end{eqnarray}
 \begin{eqnarray}
 \Pi_{1v}^{\alpha} ( q , i \omega ) & =&
 - \frac{f_0}{L^2}\sum_{k,q'} \Theta^{\alpha} ( k ) 
 \frac{ \theta( -\xi_{k+q'}^{\alpha}) 
  {\rm sgn} ( \xi^{\alpha}_k )
   \theta(-\xi_{k}^{\alpha} \xi_{k+q}^{\alpha})     }{ 
  [ \xi_{k+q'}^{\alpha} -\xi_{k+q+q'}^{\alpha}+i\omega ]
  [\xi_{k}^{\alpha}-\xi_{k+q}^{\alpha}+i\omega ] }
  \nonumber
  \\
  & + &
 \frac{f_0^{2}}{L^2}\sum_{k,q'} \Theta^{\alpha} ( k )
 \int\limits_{0}^{\infty}\mathrm{d}\omega^{\prime} \,
  \frac{ S_{\rm RPA}(q',\omega^{\prime} )}{
  [ \xi_{k}^{\alpha}-\xi_{k+q}^{\alpha}+i\omega ] 
  [ \xi_{k+q'}^{\alpha}-\xi_{k+q'+q}^{\alpha}+ i\omega ] }
 \nonumber
  \\
 & & \times 
 \bigg[
   \frac{\theta(  - \xi^{\alpha}_k \xi_{k+q'}^{\alpha} ) }{ 
  | \xi_{k}^{\alpha} | + | \xi_{k+q'}^{\alpha}| + \omega^{\prime}}
 - 
 \frac{\theta(- \xi_{k+q}^{\alpha} \xi_{k+q^{\prime}}^{\alpha})}{ | \xi_{k+q}^{\alpha}| +| \xi_{k+q'}^{\alpha}| +\omega^{\prime} + 
 i \omega \; {\rm sgn} ( \xi^{\alpha}_{k + q^{\prime}}) }
 \bigg]
 \nonumber
 \\
 &+ &
 \left[(q,i\omega )\to(-q,-i\omega )\right] \; .
 \label{eq:Pi1v}
 \end{eqnarray}
 \end{widetext}
For linearized dispersion, $\xi^{\alpha}_k \approx \alpha v_F k$,
we have verified that
$ \Pi^{\alpha}_{ 1 s} ( q , i \omega ) + \Pi^{\alpha}_{ 1 v} ( q , i \omega ) =0$,
in agreement with the closed loop theorem \cite{Dzyaloshinskii74,Kopietz97}.
With quadratic dispersion $\xi^{\alpha}_k = \alpha v_F k + k^2 / 2m$ this cancellation
is not perfect.
Assuming that  the band curvature $1/m$ is small, we may
approximate  $S_{\rm RPA} ( q , \omega ) \approx S_{\rm RPA} ( q , \omega )|_{ 1/m =0}
=
S_{\rm TLM} ( q , \omega )$ in Eqs.~(\ref{eq:Pi1s}) and (\ref{eq:Pi1v}),
because the $m$-dependence of
$S_{\rm RPA} ( q , \omega )$ is irrelevant for the cancellation between
self-energy and vertex corrections in the limit $1/m \rightarrow 0$.
One should keep in mind, however, that this is only a good approximation
as long as the dynamic structure factor is dominated by the ZS mode,
which is the case  for $  | q | / k_F \ll F$.
We shall from now on focus on this regime.
With $S_{\rm TLM} ( q , \omega )$ given in Eq.~(\ref{eq:STLM}),
the frequency integration is trivial.
However, the resulting expression is ill-defined, because 
${ \rm Im} \Pi_1^{\alpha} ( q , \omega + i 0 )$ contains singular terms
proportional to $\delta ( \omega - \xi^{\alpha}_q )$. The fact that the direct expansion of
$\Pi_{\ast} ( q , \omega )$ in powers of the interaction generates unphysical singularities
has been noticed long time ago \cite{Holas79}. 
These singularities can be avoided
by expanding the {\it{inverse}} polarization 
 $[ \Pi^{\alpha}_{\ast} ( Q ) ]^{-1}$
in powers
of the interaction, $[ \Pi^{\alpha}_{\ast} ( Q ) ]^{-1} = [ 
\Pi^{\alpha}_{0} ( Q ) ]^{-1} - \Sigma_1^{\alpha} ( Q )$, where
to first order in the RPA interaction
$ \Sigma_1^{\alpha} ( Q ) = [\Pi_0^{\alpha} ( Q ) ]^{-2} 
\Pi_1^{\alpha} ( Q )$, with 
$\Pi_1^{\alpha} ( Q ) = 
\Pi_{1s}^{\alpha} ( Q ) + \Pi_{1v}^{\alpha} ( Q )$ given 
in Eqs.~(\ref{eq:Pi1sv1}) and (\ref{eq:Pi1vv1}). 
The point is that
according to Eq.~(\ref{eq:Seffrho})
 $\Pi^{\alpha} ( Q )$ 
can be viewed as a
bosonic single-particle Green function, which should 
never be calculated directly, because its expansion in powers of the interaction 
usually contains unphysical singularities associated with the free Green function.
On the other hand, the expansion of the 
{\it{inverse}} Green function
 (i.e., the self-energy) is  expected to be regular.
Ignoring the renormalization of the real part of the
ZS dispersion, we approximate
 \begin{equation}
 \gamma_q \approx 
 {\rm Im} \Sigma_1^{\alpha_q} ( q , \omega_q + i 0)  
  [ \Pi_0^{\alpha_q } ( q , \omega_q ) ]^2   / 
 \Pi_0^{\prime}  ( q , \omega_q )
 \; ,
 \label{eq:gamma1}
 \end{equation}
where $\alpha_q = {\rm sgn} (q)$.
For simplicity we consider only the
leading behavior of $\gamma_q$ in an expansion in powers of
$q$ and $1/m$. 
The integrations over the momenta $k$ and $q^{\prime}$ 
in Eqs.~(\ref{eq:Pi1s}) and (\ref{eq:Pi1v}) 
can then be carried out exactly.
For $ | q | / k_F \ll {\rm min} \{ F, 1 \} $ we obtain the expected result
\begin{eqnarray}
 \gamma_q \approx  
 \frac{\pi}{8} \frac{F^3}{ \sqrt{ 1 + F} [ 1 + \sqrt{1 + F} ]^4} 
 \frac{ | q |^3}{ v_F m^2}
 \; . 
 \label{eq:dampres}
\end{eqnarray}
An intensity plot of the
corresponding structure factor (\ref{eq:Lorentz}) in 
Lorentzian approximation
is shown in Fig.~\ref{fig:contour}.
 \begin{figure}[tb]    
    \centering
\psfrag {y}{$\ln(I_{\rm max})$} 
 \psfrag{x}{$\ln(|q|/k_{F})$}  
\psfrag{A}{$q>0$}  
\psfrag{B}{$q<0$}  
\psfrag{C}{$\eta=0.93$}
\psfrag{D}{$\eta=0.84$}
\vspace{4mm}
  \includegraphics[scale=0.60]{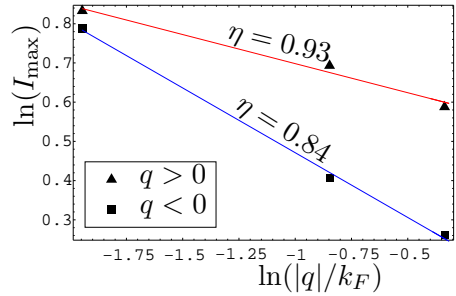}    
  \vspace{4mm}
  \caption{
Maximum  intensity $I_{\rm max} ( q )$ of the charge peak in the photoemission spectrum 
of  the quasi one-dimensional metal ${\rm{K}}_{0.3}{\rm{MoO}}_3$.
The points are data from Fig.~8 of Ref.~\cite{Gweon96}, assuming that
$q=0$ corresponds to an emission angle $\theta = 7^{\circ}$. 
The lines are fits to $I_{\rm max} (q ) \propto | q |^{-2 + 2 \eta}$.
Our theory is valid only for $  | q  | \ll k_F$, which might
explain the discrepancy between positive and negative $q$ for larger $ |q |$. 
  }
 \label{fig:photo}
 \end{figure}

Our result agrees qualitatively with Ref.~[\onlinecite{Capurro02}], which  found
$\gamma_q \propto F | q |^3 / v_F m^2$, but we disagree with
Refs.[\onlinecite{Samokhin98,Pustilnik03}], who obtained $\gamma_q \propto q^2 / m$.
Possibly  the  discrepancy with Eq.~(\ref{eq:dampres}) is related to the
non-commutativity of
the limits $q \rightarrow 0$ and $F \rightarrow 0$, which is
obvious from Eqs.~(\ref{eq:Zqrpa}) and (\ref{eq:omegaqrpa}).
Our result $\gamma_q \propto q^3$ is valid for
$ | q | / k_F \ll F$, which for fixed $| q |$ requires a minimum strength $F$ of the interaction,
but for  fixed  $F > 0$ is always satisfied for sufficiently small $q$.
In the opposite limit
$ F \ll | q | / k_F$, where most of the spectral weight 
is carried by single particle-hole excitations,
our approach also yields 
$ \gamma_q \propto q^2/m$.
The  $ q^3$-scaling of $\gamma_q$  is consistent
with a recent result by Teber \cite{Teber05}, who showed
that the attenuation rate of a coherent acoustic mode
embedded in the two-pair excitation continuum
scales as $q^3$.

$S( q , \omega )$ can be measured using
$X$-ray scattering. Unfortunately, for small $ q $ the intensity is 
very low, so that an experimental verification of Fig.~\ref{fig:contour}
seems to be difficult. However, the ZS damping
has a dramatic effect on  the shape of the
single-particle spectral function $A ( k_F + q , \omega )$ 
for $\omega \approx \omega_q$, which can be measured
via photoemission.
For the TLM (including now the spin degree of freedom) one finds 
for $\omega \rightarrow \omega_q = v_c | q |$
an algebraic singularity \cite{Meden92},
$ A ( k_F + q , \omega ) \propto  |q    (\omega - \omega_q) |^{ (  \eta -1  )/2}$,
which is a consequence of the
undamped ZS propagation.
Here $\eta$ is the anomalous dimension. 
The damping of the ZS mode
washes out this singularity. 
The sensitivity of the spectral line-shape for
$\omega \approx \omega_q$ to 
non-universal perturbations neglected in the TLM
has previously been noticed by Meden \cite{Meden99}.
In fact, from the
expression for the single-particle Green function derived via functional
bosonization \cite{Kopietz96,Kopietz97}  we estimate that 
 the  photoemission intensity exhibits 
 for $ | \omega - \omega_q | \ll 
\gamma_q$ a finite \cite{Orgad01} maximum,
$I_{\rm max}  (q ) \propto  |q  \gamma_q |^{  (\eta -1)/2 }
 \propto |q|^{ -2 + 2 \eta}$. 
In Fig.~\ref{fig:photo} we compare this prediction
with  photoemission data~\cite{Gweon96} for the ``blue bronze''
 ${\rm{K}}_{0.3}{\rm{MoO}}_3$.
The resulting value of $\eta$ is consistent with
previous estimates \cite{Gweon96} between $0.7$ and $0.9$
for this material.
However, the data \cite{Gweon96}
were taken at  $T=300K$, so that deviations from our $T=0$ theory
are expected \cite{Orgad01}. Refined photoemission data
probing the  $T=0$ regime of a Luttinger liquid would be useful.

\section{Summary and conclusions}
\label{sec:conclusions}

In summary,  using a standard diagrammatic approach
we have calculated   the  damping $\gamma_q$ of the collective charge mode
(zero sound) in a Luttinger liquid due to the non-linearity in the energy dispersion.
Our result $\gamma_q \propto | q|^3  / v_F m^2$ suggests that ZS is indeed a well defined
elementary excitation in a Luttinger liquid. We have pointed out  that 
the photoemission line-shape close to the charge peak is very sensitive
to the ZS damping and have made a  prediction
for the height of the charge peak which agrees with experiment \cite{Gweon96}.
ZS damping is also crucial to
understand  Coulomb drag experiments 
in quantum wires~\cite{Pustilnik03}. 

We emphasize that our result is based
on the expansion of the inverse irreducible polarization
to first order in the RPA interaction, using standard
many-body perturbation theory. For 
Fermi systems in three dimensions
this approach has been quite successful \cite{Holas79}, and we have
shown here that also in one dimension the perturbative calculation
of the ZS damping 
does not give rise to any  singularities, at least to first order in an expansion
in powers of the RPA interaction.
However, according to Ref.~[\onlinecite{Pustilnik06}]
higher order corrections in the {\it{bare}} interaction
(not contained in our expansion to first order in the RPA interaction)
generate new singularities which, when properly re-summed,
lead to a much larger damping, $\gamma_q \propto q^2/m$.
We suspect that the re-summation procedure proposed in
Ref.~[\onlinecite{Pustilnik06}] is unreliable, because it
does not properly take into account the closed loop theorem 
\cite{Dzyaloshinskii74,Kopietz97}. 
We are currently investigating this problem using 
a functional renormalization group approach \cite{Pirooznia07,Schuetz05} with momentum
transfer cutoff, where corrections to the RPA can be calculated systematically even
in the presence of infrared singularities. 

Finally, let us  emphasize that, in contrast to
expansions using  conventional bosonization  \cite{Samokhin98,Teber06,Aristov07},
our approach is not based on a direct expansion of the dynamic structure factor
in powers of $1/m$. This is obvious from the fact that in the
non-interacting limit we recover the exact dynamic structure factor
of the free Fermi gas with quadratic energy dispersion. 
Our approach can be formally  justified with the help of 
functional bosonization \cite{Kopietz97,Pirooznia07},
where the vertices of the effective bosonized theory
are given by symmetrized closed fermion loops
and the effective small parameter which controls  a perturbative expansion
is proportional to the combination $Fq_c / k_F$, where the range $q_c$ of the
interaction in momentum space must be small compared with $k_F = m v_F$.

\section*{ACKNOWLEDGMENTS}

We thank I. Affleck, V. Meden, K. Sch\"{o}nhammer and S. Teber for
discussions, and  R. Claessen and  G.-H. Gweon for their 
comments about 
photoemission experiments.

\end{document}